\documentclass[12pt,draft]{revtex4}

\usepackage{amsmath}
\usepackage{amsfonts}
\usepackage{latexsym}
\usepackage{amssymb} 
\usepackage{verbatim}
\usepackage{graphicx}

\newtheorem{conjecture}{Conjecture}[section]

\begin{document}

\markboth{K.~Nakamura}
{
  Gauge-invariant variables in general-relativistic perturbations
}

%
%

\title{
  Gauge-invariant variables in general-relativistic perturbations\\
  --- globalization and zero-mode problem ---
}

\author{
  Kouji Nakamura\footnote{e-mail:kouji.nakamura@nao.ac.jp}
}

\address{
  TAMA Project, Optical and Infrared Astronomy Division,\\
  National Astronomical Observatory of Japan,\\
  Osawa 2-21-1, Mitaka 181-8588, Japan
}


\date{March 31, 2012}

\begin{abstract}
  An outline of a proof of the local decomposition of linear
  metric perturbations into gauge-invariant and gauge-variant
  parts on an arbitrary background spacetime is briefly
  explained. 
  We explicitly construct the gauge-invariant and gauge-variant
  parts of the linear metric perturbations based on some
  assumptions.
  We also point out the zero-mode problem is an essential
  problem to globalize of this decomposition of linear metric
  perturbations.
  The resolution of this zero-mode problem implies the
  possibility of the development of the higher-order
  gauge-invariant perturbation theory on an arbitrary background
  spacetime in a global sense.
\end{abstract}


\maketitle



{\it 1. Introduction ---} Higher-order general-relativistic
perturbation theory is one of topical subject in recent general
relativity. 
As well-known, general relativity is based on the concept of
general covariance. 
Due to this general covariance, the ``gauge degree of freedom'', 
which is unphysical degree of freedom of perturbations, arises
in general-relativistic perturbations.
To obtain physical results, we have to fix this gauge degrees
of freedom or to extract some invariant quantities of
perturbations.
This situation becomes more complicated in higher-order
perturbation theory.
Therefore, it is worthwhile to investigate higher-order
gauge-invariant perturbation theory from a general point of
view.


According to this motivation, in Ref.~\cite{kouchan-gauge-inv},
we proposed a procedure to find gauge-invariant variables for
higher-order perturbations on an arbitrary background spacetime.
This proposal is based on the single assumption (Conjecture
\ref{conjecture:decomposition-conjecture} in this article).
Under this assumption, we summarize some formulae for the
second-order perturbations of the curvatures and energy-momentum
tensor for matter fields~\cite{kouchan-second,kouchan-second-cosmo-matter}.
Confirming this assumption in cosmological perturbations, the
second-order gauge-invariant cosmological perturbation theory
was developed~\cite{kouchan-cosmo-second,K.Nakamura:2010yg}.
Through these works, we find that our general framework of
higher-order gauge-invariant perturbation theory is well-defined  
except for the above assumption.
Therefore, we proposed the above assumption as a conjecture in
Ref.~\cite{kouchan-second-cosmo-matter}.
We also proposed a brief outline of a proof of this 
conjecture~\cite{kouchan-decomp-letter-version,K.Nakamura-2011-full-paper}.


However, in the outline of a proof in
Ref.~\cite{K.Nakamura-2011-full-paper}, special modes of 
perturbations are not included in our considerations.
We called these special modes as {\it zero modes} in
Ref.~\cite{K.Nakamura-2011-full-paper}.
Through the above proposal of our outline of a proof, we also
pointed out that the zero modes may appear in perturbation
theories on an arbitrary background spacetime. 
We called issues concerning about these zero modes as 
{\it zero-mode problem} in
Ref.~\cite{K.Nakamura-2011-full-paper}. 
At least in the current status, this zero-mode probelm is not
resolved, yet.
However, we expect that the zero modes play important roles in
some situations.
The purpose of this article is not to resolve this
zero-mode problem, but to point out an important role of the
zero modes, which is related to the globalization of
perturbations.




{\it 2. Perturbations in general relativity ---}
The notion of ``gauge'' in general relativity arise in the
theory due to the general covariance.
There are two kinds of ``gauges'' in general relativity.
These two ``gauges'' are called as the first- and the
second-kind gauges, respectively.
The distinction of these two different notion of ``gauges'' is
an important premise of our arguments.
{\it The first-kind gauge} is a coordinate system on a single
manifold ${\cal M}$.
The coordinate transformation is also called {\it gauge
  transformation of the first kind} in general relativity.
On the other hand, {\it the second-kind gauge} appears in
perturbation theories in any theory with general covariance. 
In perturbation theories, we always treat two spacetime
manifolds.
One is the physical spacetime ${\cal M}$ which is our nature
itself and we want to clarify the properties of ${\cal M}$
through perturbations.
Another is the background spacetime ${\cal M}_{0}$ which has
nothing to do with our nature but is prepared by hand for
perturbative analyses.
{\it The gauge choice of the second kind} is the point
identification map ${\cal X}$ $:$ 
${\cal M}_{0}\mapsto{\cal M}$.
We have to note that the correspondence ${\cal X}$ between
points on ${\cal M}_{0}$ and ${\cal M}$ is not unique in the 
perturbation theory with general covariance, i.e., we have no
guiding principle to choose the identification map ${\cal X}$.
Actually, as a gauge choice of the second kind, we may choose a
different point identification map ${\cal Y}$ from ${\cal X}$. 
This implies that there is degree of freedom in the gauge choice
of the second kind.
This is {\it the gauge degree of freedom of the second kind} in
general-relativistic perturbations.
{\it The gauge transformation of the second kind} is understood
as a change ${\cal X}\rightarrow{\cal Y}$ of the identification
map.


To define perturbations of an arbitrary tensor field $\bar{Q}$,
we have to compare $\bar{Q}$ on the physical spacetime 
${\cal M}_{\lambda}$ with $Q_{0}$ on the background spacetime
${\cal M}_{0}$ through the introduction of the above second-kind 
gauge choice ${\cal X}_{\lambda}$ $:$ ${\cal M}_{0}$
$\rightarrow$ ${\cal M}_{\lambda}$.
The pull-back ${\cal X}_{\lambda}^{*}$, which is induced by the
map ${\cal X}_{\lambda}$, maps a tensor field $\bar{Q}$ on 
${\cal M}_{\lambda}$ to a tensor field 
${\cal X}_{\lambda}^{*}\bar{Q}$ on ${\cal M}_{0}$. 
Once the definition of the pull-back of the gauge choice 
${\cal X}_{\lambda}$ is given, the perturbations of a tensor
field $\bar{Q}$ under the gauge choice ${\cal X}_{\lambda}$ are
simply defined by the evaluation of the Taylor expansion at 
${\cal M}_{0}$:
\begin{equation}
  \label{eq:Bruni-35}
  {}^{\cal X}\!Q
  :=
  \left.{\cal X}^{*}_{\lambda}\bar{Q}_{\lambda}\right|_{{\cal M}_{0}} 
  =
  Q_{0}
  + \lambda {}^{(1)}_{\;\cal X}\!Q
  + \frac{1}{2} \lambda^{2} {}^{(2)}_{\;\cal X}\!Q
  + O(\lambda^{3}),
\end{equation}
where ${}^{(1)}_{\;\cal X}\!Q$ and ${}^{(2)}_{\;\cal X}\!Q$ are
the first- and the second-order perturbations of $\bar{Q}$,
respectively.


When we have two different gauge choices ${\cal X}_{\lambda}$
and ${\cal Y}_{\lambda}$, we have two different representations
of the perturbative expansion (\ref{eq:Bruni-35}).
Although these two representations are different from each
other, these should be equivalent because of general
covariance.
This equivalence is guaranteed by the 
{\it gauge-transformation rules} between these two gauge
choices.
The change of the gauge choice from ${\cal X}_{\lambda}$ to
${\cal Y}_{\lambda}$ is represented by the diffeomorphism 
$\Phi_{\lambda}:=({\cal X}_{\lambda})^{-1}\circ{\cal Y}_{\lambda}$.
This diffeomorphism $\Phi_{\lambda}$ is the map $\Phi_{\lambda}$
$:$ ${\cal M}_{0}$ $\rightarrow$ ${\cal M}_{0}$ for each value
of $\lambda\in{\mathbb R}$ and does change the point
identification.
The gauge transformation $\Phi_{\lambda}$ induces a pull-back
from the representation ${}^{\cal X}Q_{\lambda}$ in the gauge
choice ${\cal X}_{\lambda}$ to the representation 
${}^{\cal Y}Q_{\lambda}$ in the gauge choice 
${\cal Y}_{\lambda}$ by 
${}^{\cal Y}Q_{\lambda}=\Phi^{*}_{\lambda} {}^{\cal X}Q_{\lambda}$.
According to generic arguments concerning the Taylor expansion
of the pull-back of tensor fields on the same
manifold~\cite{K.Nakamura:2010yg}, we obtain the
order-by-order gauge-transformation rules for the perturbative
variables ${}^{(1)}Q$ and ${}^{(2)}Q$ as
\begin{eqnarray}
  \label{eq:Bruni-47-one}
  {}^{(1)}_{\;{\cal Y}}\!Q - {}^{(1)}_{\;{\cal X}}\!Q = 
  {\pounds}_{\xi_{(1)}}Q_{0}, \quad
  {}^{(2)}_{\;\cal Y}\!Q - {}^{(2)}_{\;\cal X}\!Q = 
  2 {\pounds}_{\xi_{(1)}} {}^{(1)}_{\;\cal X}\!Q 
  +\left\{{\pounds}_{\xi_{(2)}}+{\pounds}_{\xi_{(1)}}^{2}\right\} Q_{0},
\end{eqnarray}
where $\xi_{(1)}^{a}$ and $\xi_{(2)}^{a}$ are the generators of
the gauge transformation $\Phi_{\lambda}$.


The notion of gauge invariance considered in this article is the   
{\it order-by-order gauge invariance} proposed in
Ref.~\cite{kouchan-second-cosmo-matter}. 
We call the $k$th-order perturbation ${}^{(k)}_{{\cal X}}\!Q$ is
gauge invariant iff ${}^{(k)}_{\;\cal X}\!Q = {}^{(k)}_{\;\cal Y}\!Q$
for any gauge choice ${\cal X}_{\lambda}$ and
${\cal Y}_{\lambda}$. 
Through this concept of the order-by-order gauge invariance, we
can develop the gauge-invariant perturbation theory.




{\it 3. Construction of gauge-invariant variables ---}
To construct gauge-invariant variables, we first consider the
metric perturbation.
The metric $\bar{g}_{ab}$ on ${\cal M}_{\lambda}$, which is
pulled back to ${\cal M}_{0}$ using a gauge choice 
${\cal X}_{\lambda}$, is expanded as Eq.~(\ref{eq:Bruni-35}): 
\begin{eqnarray}
  \label{eq:metric-expansion}
  {\cal X}^{*}_{\lambda}\bar{g}_{ab}
  =
  g_{ab}
  + \lambda {}_{{\cal X}}\!h_{ab}
  + \frac{1}{2} \lambda^{2}{}_{{\cal X}}\!l_{ab}
  + O^{3}(\lambda),
\end{eqnarray}
where $g_{ab}$ is the metric on ${\cal M}_{0}$.
Although this expansion depends entirely on the gauge choice
${\cal X}_{\lambda}$, henceforth, we do not explicitly express
the index of the gauge choice ${\cal X}_{\lambda}$ if there is
no possibility of confusion. 
Through these setup, in Ref.~\cite{kouchan-gauge-inv}, we 
proposed a procedure to construct gauge-invariant variables for
higher-order perturbations.
Our starting point is the following conjecture for $h_{ab}$:
\begin{conjecture}
  \label{conjecture:decomposition-conjecture}
  If there is a symmetric tensor field $h_{ab}$ of the second
  rank, whose gauge transformation rule is 
  ${}_{{\cal Y}}\!h_{ab}$ $-$ ${}_{{\cal X}}\!h_{ab}$ $=$
  ${\pounds}_{\xi_{(1)}}g_{ab}$, then there exist a tensor field
  ${\cal H}_{ab}$ and a vector field $X^{a}$ such that $h_{ab}$
  is decomposed as $h_{ab}$ $=:$ ${\cal H}_{ab}$ $+$
  ${\pounds}_{X}g_{ab}$, where ${\cal H}_{ab}$ and $X^{a}$ are
  transformed as ${}_{{\cal Y}}\!{\cal H}_{ab}$ $-$ 
  ${}_{{\cal X}}\!{\cal H}_{ab}$ $=$ $0$, 
  ${}_{\quad{\cal Y}}\!X^{a}$ $-$ ${}_{{\cal X}}\!X^{a}$ $=$
  $\xi^{a}_{(1)}$ under the gauge transformation
  (\ref{eq:Bruni-47-one}), respectively.
\end{conjecture}
In this conjecture, ${\cal H}_{ab}$ and $X^{a}$ are 
{\it gauge-invariant} and {\it gauge-variant} parts of the
perturbation $h_{ab}$.
In the case of the perturbation theory on an arbitrary
background spacetime, this conjecture is a highly non-trivial
statement due to the non-trivial curvature of the background
spacetime, though its inverse statement is trivial.


Based on Conjecture \ref{conjecture:decomposition-conjecture},
we can decompose the second-order metric perturbation $l_{ab}$
as~\cite{kouchan-gauge-inv}
\begin{eqnarray}
  \label{eq:H-ab-in-gauge-X-def-second-1}
  l_{ab}
  =:
  {\cal L}_{ab} + 2 {\pounds}_{X} h_{ab}
  + \left(
      {\pounds}_{Y}
    - {\pounds}_{X}^{2} 
  \right)
  g_{ab},
\end{eqnarray}
where ${\cal L}_{ab}$ is gauge-invariant part of the
second-order metric perturbation $l_{ab}$ and $Y^{a}$ is the
gauge-variant part of second order whose gauge-transformation
rule is given by ${}_{{\cal Y}}\!Y^{a}$ $-$ 
${}_{{\cal X}}\!Y^{a}$ $=$ $\xi_{(2)}^{a}$ $+$ $[\xi_{(1)},X]^{a}$. 
Furthermore, using the first- and second-order gauge-variant
parts, $X^{a}$ and $Y^{a}$, of the metric perturbations, the
gauge-invariant variables for an arbitrary tensor field $Q$
are given by 
\begin{eqnarray}
  \label{eq:matter-gauge-inv-defs}
  {}^{(1)}\!{\cal Q} := {}^{(1)}\!Q - {\pounds}_{X}Q_{0}
  , \quad 
  {}^{(2)}\!{\cal Q} &:=& {}^{(2)}\!Q - 2 {\pounds}_{X} {}^{(1)}Q 
  - \left\{ {\pounds}_{Y} - {\pounds}_{X}^{2} \right\} Q_{0}
  .
\end{eqnarray}


In Ref.~\cite{kouchan-gauge-inv}, we extended this construction
to the third-order perturbations and we have already confirmed
that this construction is valid in the fourth-order
perturbations~\cite{kouchan-arbitrary-order-progress}.




{\it 4. An outline of a proof of Conjecture
  \ref{conjecture:decomposition-conjecture} ---}
To give an outline of a proof of Conjecture
\ref{conjecture:decomposition-conjecture} for an arbitrary
background spacetime, we assume that background spacetimes admit
ADM decomposition.
Therefore, the background spacetime ${\cal M}_{0}$ considered
here is $n+1$-dimensional spacetime which is described by the
direct product ${\mathbb R}\times\Sigma$.
Here, ${\mathbb R}$ is a time direction and $\Sigma$ is the
spacelike hypersurface ($\dim\Sigma = n$) embedded in
${\cal M}_{0}$.
This means that ${\cal M}_{0}$ is foliated by the one-parameter
family of spacelike hypersurface $\Sigma(t)$, where
$t\in{\mathbb R}$ is a time function.
Then, the metric on ${\cal M}_{0}$ is described by the ADM
decomposition
\begin{eqnarray}
  \label{eq:gdb-decomp-dd-minus-main}
  g_{ab} &=& - \alpha^{2} (dt)_{a} (dt)_{b}
  + q_{ij}
  (dx^{i} + \beta^{i}dt)_{a}
  (dx^{j} + \beta^{j}dt)_{b},
\end{eqnarray}
where $\alpha$ is the lapse function, $\beta^{i}$ is the
shift vector, and $q_{ij}$ is the metric on $\Sigma(t)$.


Since the ADM decomposition
(\ref{eq:gdb-decomp-dd-minus-main}) is a local one, we
may regard that our arguments are restricted to that for a
single patch in ${\cal M}_{0}$ which is covered by 
the metric (\ref{eq:gdb-decomp-dd-minus-main}).
Therefore, we regard $\Sigma$ as this single patch of a
spacelike hypersurface in ${\cal M}_{0}$.
Further, we may change the region which is covered by the metric
(\ref{eq:gdb-decomp-dd-minus-main}) through the choice of
the lapse function $\alpha$ and the shift vector $\beta^{i}$.
The choice of $\alpha$ and $\beta^{i}$ is regarded as the
first-kind gauge choice, which have nothing to do with the
second-kind gauge.
Since we may regard that the representation 
(\ref{eq:gdb-decomp-dd-minus-main}) of the background
metric is that on a single patch in ${\cal M}_{0}$, in general
situation, each $\Sigma$ may have its boundaries $\partial\Sigma$.


To prove Conjecture
\ref{conjecture:decomposition-conjecture}, we consider the
components of the metric $h_{ab}$ as $h_{ab}$ $=$
$h_{tt}(dt)_{a}(dt)_{b}$ $+$ $2h_{ti}(dt)_{(a}(dx^{i})_{b)}$ $+$ 
$h_{ij}(dx^{i})_{a}(dx^{j})_{b}$.
The gauge-transformation rules for the components $\{h_{tt}$,
$h_{ti}$, $h_{ij}\}$ are derived from ${}_{{\cal Y}}\!h_{ab}$ $-$
${}_{{\cal X}}\!h_{ab}$ $=$ ${\pounds}_{\xi_{(1)}}g_{ab}$ with
$\xi_{(1)a}$ $=$ $\xi_{t}(dt)_{a}$ $+$ $\xi_{i}(dx^{i})_{a}$.
Inspecting these gauge-transformation rules, we explicitly
construct gauge-invariant and gauge-variant variables.


Our strategy for the proof is as
follows~\cite{kouchan-decomp-letter-version,K.Nakamura-2011-full-paper}: 
we first assume that the existence of the variables $X_{t}$ and
$X_{i}$ whose gauge-transformation rules are given by 
${}_{{\cal Y}}X_{t}$ $-$ ${}_{{\cal X}}X_{t}$ $=$ $\xi_{t}$ and 
${}_{{\cal Y}}X_{i}$ $-$ ${}_{{\cal X}}X_{i}$ $=$ $\xi_{i}$,
respectively. 
This assumption is confirmed through the explicit construction
of the gauge-variant part of the linear-order metric
perturbation below. 
Further, inspecting gauge-transformation rules for the
components $\{h_{tt}$, $h_{ti}$, $h_{ij}\}$, we define the
symmetric tensor field $\hat{H}_{ab}$ whose components are given
by
\begin{eqnarray}
  \hat{H}_{tt}
  &:=&
  h_{tt} 
  + \frac{2}{\alpha}\left(
    \partial_{t}\alpha 
    + \beta^{i}D_{i}\alpha 
    - \beta^{j}\beta^{i}K_{ij}
  \right) X_{t}
  \nonumber\\
  &&
  + \frac{2}{\alpha} \left(
    \beta^{i}\beta^{k}\beta^{j} K_{kj}
    - \beta^{i} \partial_{t}\alpha
    + \alpha q^{ij} \partial_{t}\beta_{j}
  \right.
  \nonumber\\
  && \quad\quad\quad
  \left.
    + \alpha^{2} D^{i}\alpha 
    - \alpha \beta^{k} D^{i} \beta_{k}
    - \beta^{i} \beta^{j} D_{j}\alpha 
  \right)X_{i}
  \label{eq:hatHtt-def-generic}
  , \\
  \hat{H}_{ti}
  &:=&
  h_{ti}
  + \frac{2}{\alpha} \left(
    D_{i}\alpha 
    - \beta^{j}K_{ij}
  \right) X_{t}
  + \frac{2}{\alpha} M_{i}^{\;\;j} X_{j}
  \label{eq:hatHti-def-generic}
  , \\
  \hat{H}_{ij}
  &:=&
  h_{ij}
  - \frac{2}{\alpha} K_{ij} X_{t}
  + \frac{2}{\alpha} \beta^{k} K_{ij} X_{k}
  \label{eq:hatHij-def-generic}
  ,
\end{eqnarray}
where $M_{i}^{\;\;j}$ is defined by $M_{i}^{\;\;j}$ $:=$ $-$
$\alpha^{2}K^{j}_{\;\;i}$ $+$ $\beta^{j}\beta^{k} K_{ki}$ $-$
$\beta^{j}D_{i}\alpha$ $+$ $\alpha D_{i}\beta^{j}$. 
Here, $K_{ij}$ is the components of the extrinsic curvature of 
$\Sigma$ in ${\cal M}_{0}$ and $D_{i}$ is the covariant
derivative associate with the metric $q_{ij}$ ($D_{i}q_{jk}=0$). 
The extrinsic curvature $K_{ij}$ is related to the time
derivative of the metric $q_{ij}$ by $K_{ij}$ $=$ $-$
$(1/2\alpha)$
$\left[\partial_{t}q_{ij}-D_{i}\beta_{j}-D_{j}\beta_{i}\right]$.
The gauge transformation rules for the components of
$\hat{H}_{ab}$ are given by  
\begin{eqnarray}
  {}_{{\cal Y}}\!\hat{H}_{tt}
  -
  {}_{{\cal X}}\!\hat{H}_{tt}
  =
  2 \partial_{t}\xi_{t}
  , \quad
  {}_{{\cal Y}}\!\hat{H}_{ti}
  -
  {}_{{\cal X}}\!\hat{H}_{ti}
  =
  \partial_{t}\xi_{i}
  + D_{i}\xi_{t}
  , \quad
  {}_{{\cal Y}}\!\hat{H}_{ij}
  -
  {}_{{\cal X}}\!\hat{H}_{ij}
  =
  2 D_{(i}\xi_{j)}
  .
  \label{eq:gauge-trans-hatHij}
\end{eqnarray}


Since the components $\hat{H}_{it}$ and $\hat{H}_{ij}$ are
regarded as components of a vector and a symmetric tensor on
$\Sigma$, respectively, we may apply the following
decomposition~\cite{J.W.York-1973-1974} to $\hat{H}_{ti}$ and
$\hat{H}_{ij}$: 
\begin{eqnarray}
  \label{eq:K.Nakamura-2010-2-generic-4-7}
  \hat{H}_{ti} &=:& D_{i}h_{(VL)} + h_{(V)i}, \quad D^{i}h_{(V)i} = 0,
  \\
  \label{eq:K.Nakamura-2010-2-generic-4-8}
  \hat{H}_{ij} &=:& \frac{1}{n} q_{ij} h_{(L)} + h_{(T)ij}, \quad
  q^{ij} h_{(T)ij} = 0,
  \\
  \label{eq:K.Nakamura-2010-2-generic-4-9}
  h_{(T)ij} &=:& \left(Lh_{(TV)}\right)_{ij} + h_{(TT)ij},
  \quad
  D^{i}h_{(TT)ij} = 0,
\end{eqnarray}
where $(Lh_{(TV)})_{ij}$ is defined by $(Lh_{(TV)})_{ij}$ $:=$
$D_{i}h_{(TV)j}$ $+$ $D_{j}h_{(TV)i}$ $-$
$\frac{2}{n}q_{ij}D^{l}h_{(TV)l}$.  
Equations (\ref{eq:gauge-trans-hatHij}) give the
gauge-transformation rules for the variables $h_{(VL)}$,
$h_{(V)i}$, $h_{(L)}$, $h_{(T)ij}$, $h_{(TV)i}$, and $h_{(TT)ij}$.
From these gauge-transformation rules, we easily find the
explicit form of the variables $X_{t}$ and $X_{i}$ as
follows~\cite{kouchan-decomp-letter-version,K.Nakamura-2011-full-paper}: 
\begin{eqnarray}
  X_{i} := h_{(TV)i}
  , \quad
  X_{t}
  :=
  h_{(VL)}
  - \Delta^{-1}D^{k}\partial_{t}h_{(TV)k}
  .
  \label{eq:K.Nakamura-2010-note-B-52}
\end{eqnarray}
This is the most non-trivial part in our outline of a proof of
Conjecture \ref{conjecture:decomposition-conjecture}.
Further, we have to note that {\it in the derivation of
  Eqs.~(\ref{eq:K.Nakamura-2010-note-B-52}), we assume that the
  existence of the Green functions $\Delta^{-1}$ and 
  $({\cal D}^{ij})^{-1}$ of the Laplacian $\Delta:=D^{i}D_{i}$
  and the elliptic derivative operator ${\cal D}^{ij}$}
defined by ${\cal D}^{ij}$ $:=$ $q^{ij}\Delta$ $+$ 
$\left(1-\frac{2}{n}\right)D^{i}D^{j}$ $+$ $R^{ij}$.
Here, $R^{ij}$ is the Ricci curvature on $\Sigma$.
In other words, {\it we have ignored the perturbative modes
  which belong to the kernel of the derivative operators
  $\Delta$ and ${\cal
    D}^{ij}$}~\cite{kouchan-decomp-letter-version,K.Nakamura-2011-full-paper}.  
We call these modes as {\it zero modes}.


Furthermore, we easily construct gauge-invariant variables for
the linear-order metric perturbation $h_{ab}$.
We have two scalar modes ($\Phi$ and $\Psi$), one transverse
vector mode $\nu_{i}$, one transverse-traceless tensor mode
$\chi_{ij}$.
These gauge-invariant variables are given by 
\begin{eqnarray}
  - 2 \Phi
  &:=&
  \hat{H}_{tt}
  - 2 \partial_{t}X_{t}
  ,
  \quad
  - 2 n \Psi
  :=
  h_{(L)} - 2 D^{i}X_{i}
  ,
  \nonumber\\
  \nu_{i}
  &:=&
  h_{(V)i}
  - \partial_{t}X_{i}
  + D_{i}\Delta^{-1}D^{k}\partial_{t}X_{k}
  , \quad
  \chi_{ij}
  :=
  h_{(TT)ij}
  .
  \label{eq:K.Nakamura-2010-note-B-55}
\end{eqnarray}
Moreover, we can derive the expressions of the original
components $\{h_{tt}$, $h_{ti}$, $h_{ij}\}$ of the metric
perturbation $h_{ab}$ in terms of these gauge-invariant
variables and the variables $X_{t}$ and $X_{i}$.
Then, we conclude that we may identify the components of the
gauge-invariant variables ${\cal H}_{ab}$ and the gauge-variant
variable $X_{a}$ so that ${\cal H}_{tt}$ $:=$ $-2\Phi$, 
${\cal H}_{ti}$ $:=$ $\nu_{i}$, ${\cal H}_{ij}$ $:=$ 
$-2\Psi q_{ij}$ $+$ $\chi_{ij}$, $X_{a}$ $:=$ $X_{t}(dt)_{a}$
$+$ $X_{i}(dx^{i})_{a}$. 
These identifications lead to the assertion of Conjecture
\ref{conjecture:decomposition-conjecture}.




{\it 5. Zero-mode problem and the globalization of
  gauge-invariant variables ---}
In the above outline of a proof of Conjecture
\ref{conjecture:decomposition-conjecture},
we concentrate only on a local region $\Sigma$ in a spacelike
hypersurface which is covered by the metric
(\ref{eq:gdb-decomp-dd-minus-main}).
This local region $\Sigma$ may have its boundaries
$\partial\Sigma$.
Furthermore, we assumed the existence of Green functions of the
elliptic derivative operators $\Delta$ or ${\cal D}^{ij}$.
Since we concentrated only on a local region $\Sigma$ of the
whole spacelike hypersurface in the above outline of a proof, we
have to discuss the globalization of our proof to the whole
region of the spacelike hypersurface in the background spacetime
${\cal M}_{0}$ if we insist that Conjecture 
\ref{conjecture:decomposition-conjecture} is true on the whole
background spacetime manifold ${\cal M}_{0}$.
In my opinion, the key of this globalization is zero modes.


As mentioned above, we define zero modes as perturbative
modes which belongs to the kernel of the elliptic derivative
operators $\Delta$ or ${\cal D}^{ij}$.
The kernel of ${\cal D}^{ij}$ also includes the (conformal)
Killing vectors. 
Therefore, we may say that zero modes are related to the
symmetries of the background spacetime.
Furthermore, we should emphasize that we have to impose
boundary conditions at boundaries $\partial\Sigma$ for the
explicit construction of the Green functions for $\Delta$ and
${\cal D}^{ij}$.
Since the operators $\Delta$ and ${\cal D}^{ij}$ are elliptic,
the change of the boundary conditions at $\partial\Sigma$ is
adjusted by functions which belong to the kernel of the
operators $\Delta$ and ${\cal D}^{ij}$, i.e., zero modes.
Thus, we may say that the informations for the boundary
conditions for the Green functions $\Delta^{-1}$ and
$\left({\cal D}^{ij}\right)^{-1}$ are also included in the zero
modes.
These modes should be separately treated in different manner.
We call the issue concerning about treatments of these zero
modes as the {\it zero-mode problem}.
This problem is a remaining problem in our general framework on
higher-order general-relativistic gauge-invariant perturbation
theory.


Now, we comment on the relation between the globalization of
our outline of a proof and the zero-mode problem.
If we want to consider the global behaviors of perturbations, we
have to consider the globalization of the definition of the
gauge-invariant and gauge-variant variables to the whole region
of a spacelike hypersurface in ${\cal M}_{0}$.
To do this, we should consider different patches which cover the
region outside the local region $\Sigma$.
The same arguments as above is applied to perturbative variables
on these different patches.
The key problem is how to identify the gauge-invariant and
gauge-variant variables for perturbations on these different
patches to those on $\Sigma$.
To accumplish this, the behavior of the perturbative variables
at the boundaries $\partial\Sigma$ is important.
If we impose some smoothness of the gauge-invariant and
gauge-variant variables on the whole region of a spacelike
hypersuface, we have to impose an appropriate boundary
conditions to the perturbative variables at $\partial\Sigma$ and
to match the perturbative variables at $\partial\Sigma$ with
those on the region outside $\Sigma$.
If we accumplish this matching, we may regard that the
gauge-invariant and gauge-variant variables are global variables
on the whole region of a spacelike hypersurface on $\Sigma$. 
As mentioned above, the change of the boundary conditions at
$\partial\Sigma$ is adjusted by zero modes.
Thus, zero modes is necessary to construct global
gauge-invariant and gauge-variant variables.


If we impose the Einstein equation as the field equation, this
globalization is related to the construction of a global
solution to the perturbative Einstein equation.
To construct perturbative solutions to the initial value
constraints, we have to consider the perturbative solutions
outside the local region $\Sigma$ and to match with the
perturbative solutions in $\Sigma$ at $\partial\Sigma$. 
To accumplish this matching, each solution to the initial value
constraints will be required to satisfy some appropriate
boundary conditions at the boundaries $\partial\Sigma$.
The boundary behavior of the perturbative variables is also
adjusted by the zero mode which satisfy the perturbative
Einstein equation.
If we accumplish this matching of solutions to the initial value
constraint smoothly, we can consider the time evolution of the
global perturbations following to the evolution equations in the
Einstein equation.


We also note that to ignore the zero modes is regarded as to
impose boundary conditions for the Green functions $\Delta$ or
${\cal D}^{ij}$ at the boundary $\partial\Sigma$ in some way.
There is no guarantee whether this boundary condition at
$\partial\Sigma$ is appropriate to construct global solution to
the perturbative Einstein equation or not. 
The information at $\partial\Sigma$ propagates along the
boundary of the domain of dependence of $\Sigma$ through the 
dynamics of the Einstein equation.
If the imposed boundary conditions are not appropriate to
construct global solutions, the obtained solution to
perturbative Einstein equation cannot be extend to the outside
of the domain of dependence of $\Sigma$ in general and loses its
physical relevance of the behavior at the boundary of the domain
of dependence of $\Sigma$.
In this sense, zero modes are the important to construct global
perturbative solutions.
This is the main point of this article.




{\it 6. Summary ---} 
We briefly explained our proposal of an outline of a proof 
Conjecture \ref{conjecture:decomposition-conjecture} for an
arbitrary background spacetime.
Although there will be many approaches to prove Conjecture
\ref{conjecture:decomposition-conjecture}, in this
article, we just show an outline a proof.
We also note that our arguments do not include zero modes.
The existence of zero modes is also related to the symmetries of
the background spacetime.
Furthermore, the zero modes are also important to construct
global gauge-invariant and gauge-variant variables of
perturbations and to derive global solutions to the perturbative
Einstein equations. 
To resolve this zero-mode problem, careful discussions on
domains of functions for perturbations and its boundary
conditions at $\partial\Sigma$ will be necessary. 
If we resolved this zero-mode problem, the general framework of
the general-relativistic higher-order gauge-invariant
perturbation theory will be completed and the wide applications
of this general framework will be opened.




\end{document}